\input{aipcheck}

\documentclass[
    ,final            
  ]
  {aipproc}

\usepackage{epstopdf}

\layoutstyle{6x9}

\begin{document}

\title{Simulating Quantum Magnetism with Correlated Non-Neutral Ion Plasmas}

\classification{03.67.-a; 37.10.Ty; 52.27.Gr; 52.27.Jt }

\keywords      {ion traps; Ising interaction; Penning trap; quantum simulation}

\author{John J. Bollinger}{
  address={Time and Frequency Div., NIST, Boulder, CO 80305, USA}
}

\author{Joseph W. Britton}{
  address={Time and Frequency Div., NIST, Boulder, CO 80305, USA}
}

\author{Brian C. Sawyer}{
  address={Time and Frequency Div., NIST, Boulder, CO 80305, USA}
}

\begin{abstract}
 By employing forces that depend on the internal electronic state (or
spin) of an atomic ion, the Coulomb potential energy of a strongly
coupled array of ions can be modified in a spin-dependent way to mimic
effective quantum spin Hamiltonians. Both ferromagnetic and anti-ferromagnetic
interactions can be implemented. We use simple models to explain how the effective spin interactions are engineered with trapped-ion crystals. We summarize the type of
effective spin interactions that can be readily generated, and discuss
an experimental implementation using single-plane ion crystals in a Penning trap.
\end{abstract}

\maketitle


\section{Introduction}

  Currently, many important problems in physics, for example solid-state
physics, are poorly understood because the underlying quantum mechanics
is too complex for meaningful modeling. Frequently cited examples
include correlated magnetic systems, such as spin liquids~\cite{Balents:2010}, and high-temperature (high-Tc)
superconductivity, where we are not able to calculate the phase
diagram or even know what phases exist in the Fermi-Hubbard model~\cite{Georges:1996},
a leading candidate for explaining high-Tc superconductivity.
The source of this problem is the exponential increase in the difficulty
of quantum many-body calculations as the number of bodies increases.
Considering simple two-level quantum systems or spins, the complexity
of the calculation doubles with each additional spin. This exponential
increase in complexity with the number of spins makes a direct calculation
of a general system of more than $\sim$ 30 interacting
spins intractable on current computers~\cite{Sandvik:2010}.

In an attempt to tackle this difficulty, a number of groups
are following a suggestion put forth by Feynman~\cite{Feynman:1982} that it might be possible to engineer interactions between well controlled
quantum components in order to mimic a quantum many-body system that is not
understood. Through measurement of the well controlled quantum components,
it may then be possible to obtain the solution, or at least acquire
some information about the behavior of the currently intractable quantum many-body system. Current platforms (i.e., well controlled quantum systems) being used
for quantum simulation include neutral atoms in optical lattices~\cite{Jordens:2008},
superconducting circuits~\cite{Houck:2012}, and trapped ions~[7-11]. Each quantum platform has particular strengths for engineering different types of many-body
Hamiltonians. Neutral atoms in optical lattices are well suited for
simulating many-body Hamiltonians where quantum statistics (Fermi
or Bose) are important. Crystalline arrays of trapped ions are a promising
platform for emulating quantum magnetic systems. Superconducting circuits
provide exceptional flexibility in the design of interactions between
elementary quantum building blocks.

In the next section we use a simple model to illustrate
how crystalline arrays of trapped ions can be used to emulate an interacting
system of spin-1/2 particles. Crystalline arrays of trapped ions are
strongly coupled; the potential energy due to the Coulomb interaction
is much larger than the ion thermal energy. The basic idea behind engineering
quantum magnetic interactions is to generate an internal-state (or spin) dependence of the Coulomb interaction energy, through the application of
forces that depend on an ion's internal state. In the final section we
describe our experimental set-up for implementing quantum magnetic
interactions on single-plane triangular arrays of ions stored in a
Penning trap~\cite{Britton:2012}.

\section{Quantum magnetic interactions through spin-dependent forces}


Ions have a variety of different internal states. For example, the
ground and excited electronic configurations of an ion give rise to
different internal states. Each of these configurations is typically
split into a number of different hyperfine states corresponding to
different orientations of the electronic spin and nuclear moment.
With well developed techniques from atomic physics, it is frequently possible to isolate
and control two different internal atomic levels --- a two-level system or qubit
--- with high fidelity. This two-level system acts as an effective spin-1/2~\cite{Feynman:1957} and we denote these two levels with the notation $|\uparrow\rangle, |\downarrow\rangle$.
In our experimental set-up (described later), the effective spin-1/2 is the valence electron spin-flip transition in the ground state of Be$^{+}$ in the 4.46 T magnetic
field of the Penning trap. In this case the isolated two-level system
is very close to a real spin-1/2 particle, but this does not need
to be the case. Because the distance $d$ between ions in a trapped
ion crystal is large (typically $d \sim$many micrometers), the
interaction energy between two neighboring ions in the crystal lattice
is simply the spin-independent Coulomb
interaction $e^{2}/(4\pi\epsilon_{o} d)$. For the valence-electron spin qubit discussed
here, the magnetic dipole interaction energy at a separation of $d \sim$10 $\mu$m is $\frac{\mu_{o}}{4\pi} \frac{\mu_{B}^2}{d^3} \sim 8.6\times 10^{-39}$ J $\sim h$($12$ $\mu$Hz), which is too weak to be of any consequence.

\begin{figure}
  \includegraphics[height=.62\textheight]{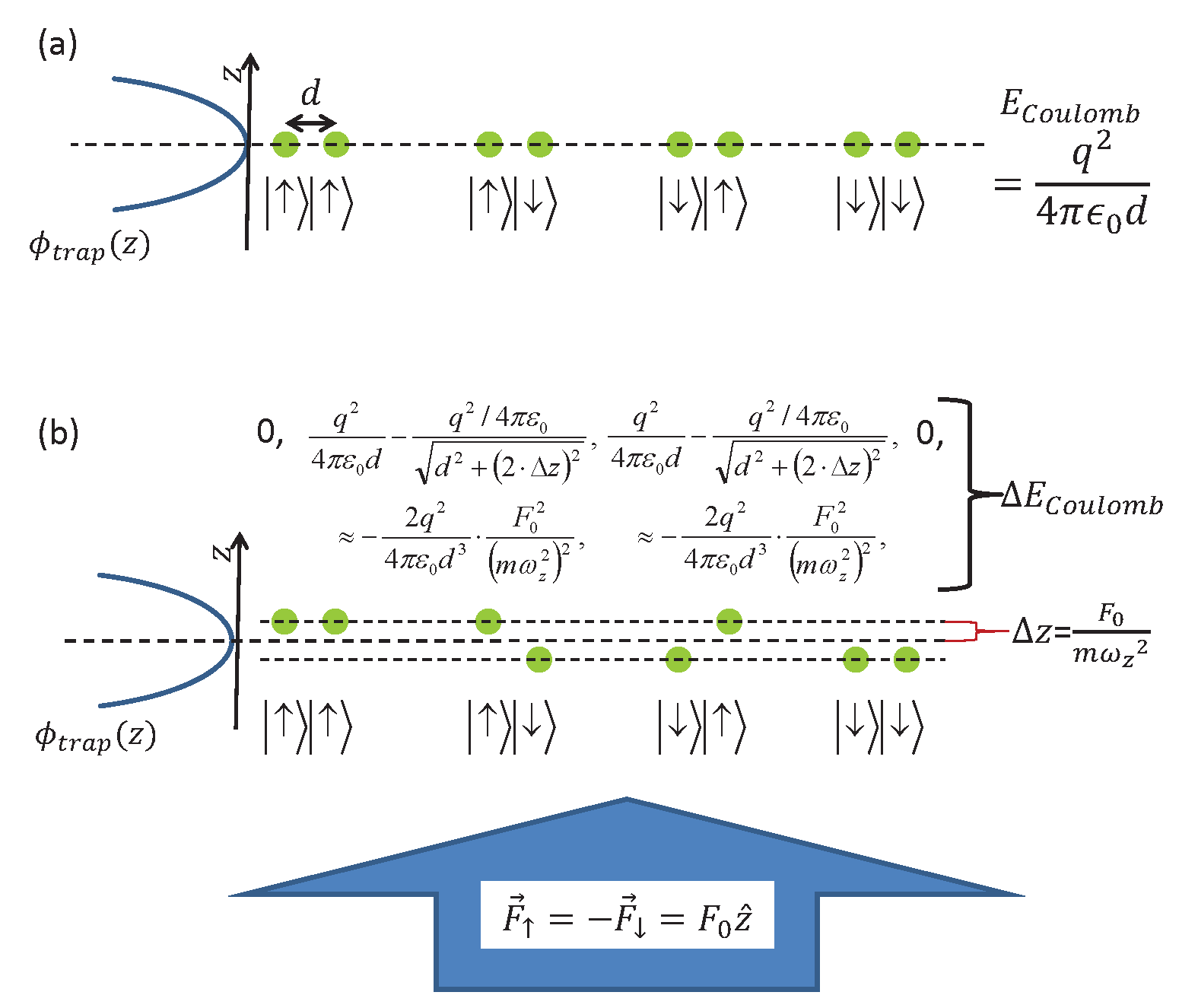}
  \caption{(a) Simple example of two trapped ions separated by a distance $d$. In a direction normal to their separation the ions reside in a harmonic potential characterized by frequency $\omega_z$. The four different spin states are shown. (b) Application of a static, spin-dependent force produces spin-dependent displacements of the ions.  If the magnitude $|\Delta z|$ of the displacements are identical for all the spin states, the change in the Coulomb potential energy looks like an effective antiferromagnetic interaction.}
\label{static force no trap}
\end{figure}

Larger spin-spin interactions with trapped ions are obtained through
the application of spin-dependent forces. A detailed description can get complicated, but the basic idea can be illustrated for the simple case of a two-ion crystal, as shown in Fig. \ref{static force no trap}.  Suppose the ions are confined by a harmonic trapping potential characterized by a frequency $\omega_z$ in the direction transverse to their separation . With two ions there are 4 different spin states shown in Fig.
\ref{static force no trap}(a) that need to be considered: two ferromagnetic states $(|\uparrow\rangle|\uparrow\rangle,\,|\downarrow\rangle|\downarrow\rangle)$ and two antiferromagnetic states $(|\uparrow\rangle|\downarrow\rangle,\,|\downarrow\rangle|\uparrow\rangle)$.  Now suppose we adiabatically turn
on a spin-dependent force in a direction transverse to the separation
of the two spins (the z direction in Fig. \ref{static force no trap}). The spin-dependent force
could be, for example, due to an applied inhomogeneous magnetic field or an optical dipole force. For simplicity we assume that
the force $\vec{F}_{\uparrow}$ on $\left|\uparrow\right\rangle $
is equal in magnitude but opposite in sign to the force $\vec{F}_{\downarrow}$
on $\left|\downarrow\right\rangle $. If, for the moment, we neglect the Coulomb repulsion between the ions, then each spin $\left|\uparrow\right\rangle (\left|\downarrow\right\rangle )$ state
is displaced from the trap center by a distance $\Delta z=F_{0}/(m\omega_{z}^2)\:\left(\Delta z=-F_{0}/(m\omega_{z}^2)\right)$. Because all ions are displaced from the trap center by the same distance, the increase in the external trapping potential energy $\Delta E_{Trap}$ is the same for all spin states. However, for the anti-ferromagnetic states $\left(\left|\uparrow\right\rangle \left|\downarrow\right\rangle ,\:\left|\downarrow\right\rangle \left|\uparrow\right\rangle \right)$,
the Coulomb interaction energy is reduced because of the increased
distance between the ions. This change in the Coulomb interaction energy
mimics an antiferromagnetic interaction between the two ion qubits. This illustrates, at its most basic level, the idea behind generating effective spin-spin interactions (and quantum gates~\cite{Leibfried:2003,Calarco:2001}) with trapped ions.

\begin{figure}
  \includegraphics[height=.45\textheight]{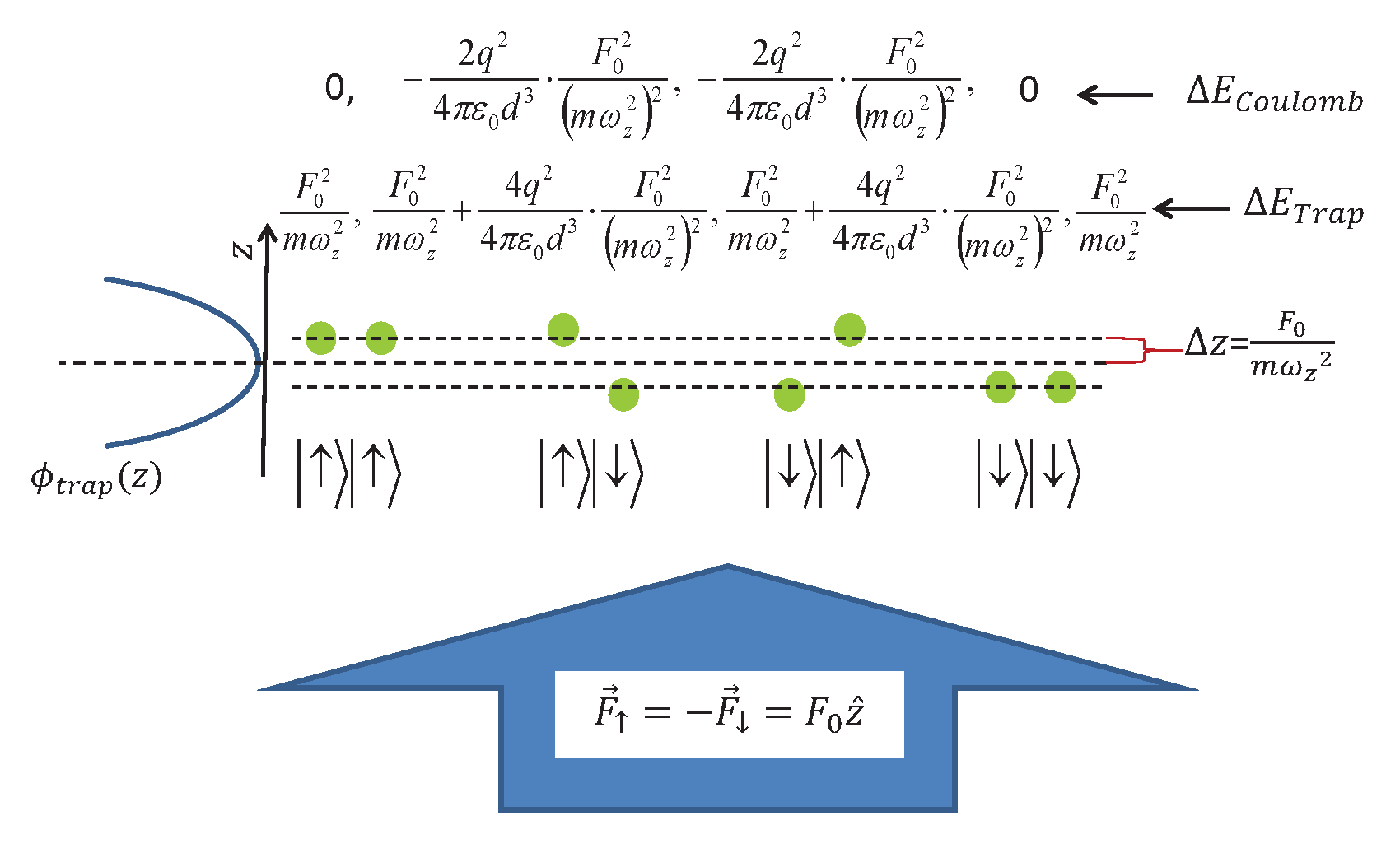}
  \caption{ For the antiferromagnetic states, the Coulomb repulsion between the ions produces a larger displacement of the ions from trap center and an increase in the external trapping potential energy relative to the ferromagnetic states.  The total change in energy mimics an ferromagnetic interaction. }
\label{static force enhance}
\end{figure}

However, an important detail has been left out of Fig. \ref{static force no trap}. In particular, if we include the Coulomb repulsion between the ions, then the displacement of an ion from the trap center depends on the spin state of the other ion. For the antiferromagnetic states, the Coulomb repulsion enhances the effect of the spin-dependent
force, producing a larger displacement of the ions from trap center
and an increase in the external trapping potential energy
relative to the ferromagnetic states. As sketched in Fig. \ref{static force enhance}, for small displacements $(\Delta z <<d)$ the increase in the trapping
potential energy is twice the decrease in the Coulomb interaction energy,
with the net result that a static, spin-dependent force actually
produces an effective ferromagnetic Ising interaction of strength
\[
H_{Ising}=J\sigma_{1}^{z}\cdot\sigma_{2}^{z},\: J\simeq-\frac{e^{2}}{4\pi \epsilon_{0} d^{3}}\cdot\frac{F_{0}^{2}}{\left(m\omega_{z}^{2}\right)^{2}}\:.
\]
An engineered ferromagnetic interaction can be used, along with a transverse field, to study a paramagnetic-to-ferromagnetic quantum phase transition~\cite{Friedenauer:2008}. However, for the triangular lattice discussed in the next section, an antiferromagnetic interaction would be particulary interesting, as it enables studies of magnetic frustration and possible spin-liquid behavior.

An antiferromagnetic interaction can be engineered through application
of an oscillating spin-dependent force $\vec{F}_{\uparrow}(t)=-\vec{F}_{\downarrow}(t)=\vec{F}_{0}\sin\mu_{R}t$.
Once again, consider slowly ramping up the amplitude of
the spin-dependent force. We are interested in the steady-state modification
to the energy of the system due to the oscillating spin-dependent
force. If the frequency of the applied spin-dependent force is greater
than the trapping frequency ($\mu_{R}>\omega_{z}$), the driven oscillation
of the ions is $180^{o}$ out of phase with the applied force,
the exact opposite of that discussed above for the $\mu_{R}=0$. For the antiferromagnetic states the Coulomb repulsion between the ions now opposes the spin-dependent force, resulting in a decrease
in the driven amplitude of the antiferromagnetic states (relative
to the ferromagnetic states). For $\mu_r > \omega_z$ the energy of the driven motion is therefore greater for the ferromagnetic states than for the antiferromagnetic states, resulting in an antiferromagnetic Ising interaction.

These simple examples illustrate the basic idea behind engineering spin-spin interactions with trapped ion crystals.  Through the application of spin-dependent forces, the total energy of the crystal can be modified in a way that depends on the internal "spin-state" of the ions, generating an effective spin-spin interaction. The sign of the interaction can be tuned by employing a sinusoidally varying spin-dependent force: for $\mu_r > \omega_z$ the effective interaction is antiferromagnetic; for $\mu_r << \omega_z$ the interaction is ferromagnetic.  More complicated time-dependent forces can generate arbitrary spin-spin interactions~\cite{Korenblit:2012}.

\section{Engineered Ising interactions on single-plane crystals}

\begin{figure}
  \includegraphics[height=.45\textheight]{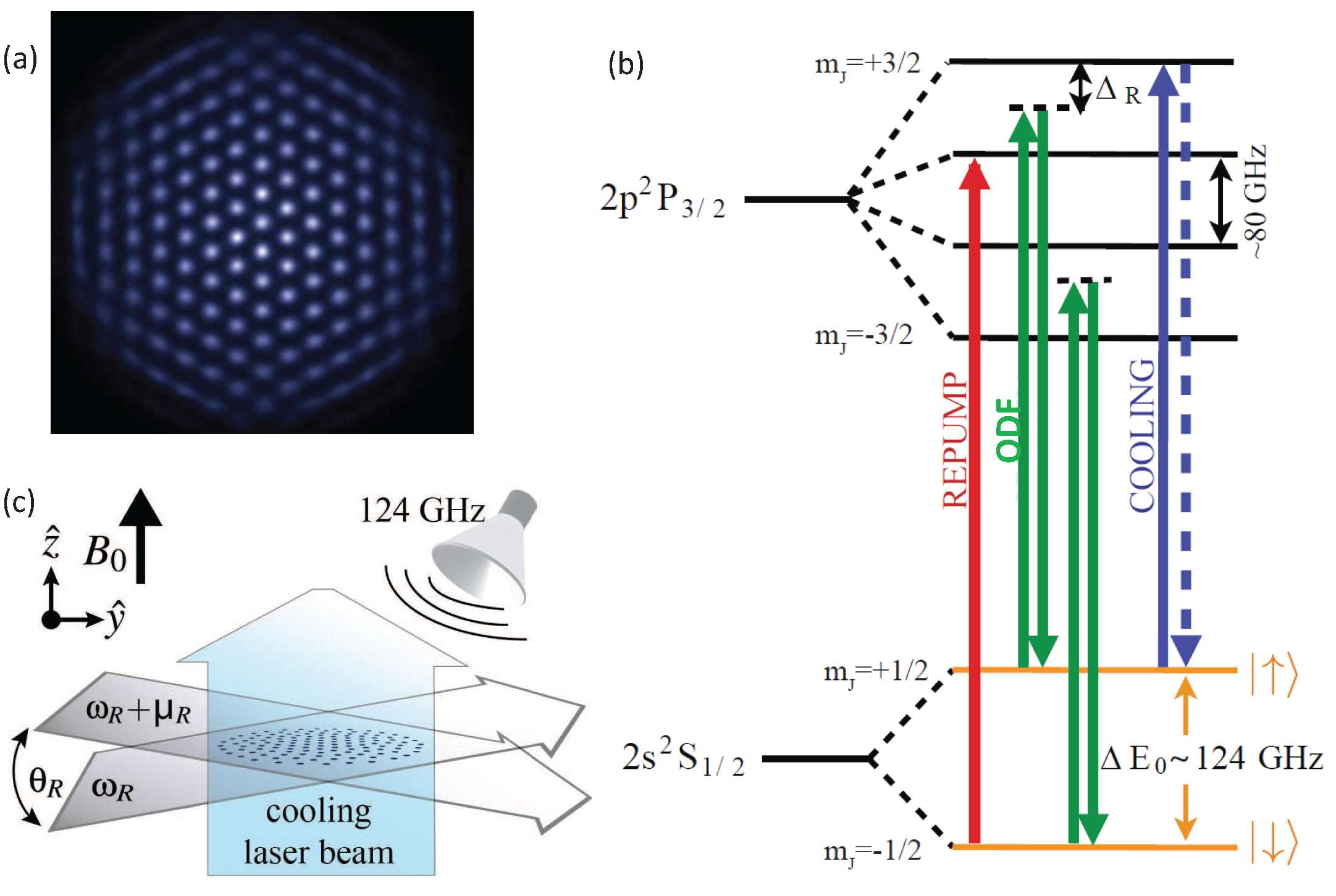}
  \caption{ (a) Top-view resonance fluorescence image showing the center region of an ion crystal captured in the ion's rest frame~\cite{Mitchell:2001}. The nearest neighbor ion spacing is $\sim$20 $\mu$m. (b) Relevant energy levels of $^9$Be$^+$ at $B_0$=4.46 T (not drawn to scale). The $^{9}$Be$^{+}$ nucleus
has spin $I=3/2$. We show only $m_I=3/2$ levels that are prepared experimentally through optical pumping. The $^2 S_{1/2} - ^2 P_{3/2}$ transition wavelength is $\sim$313 nm. The qubit splitting $\Delta E_0 \sim 124$ GHz. A low-phase-noise microwave source at 124 GHz provides full global control over spins. (c) The optical dipole force (ODF) interaction is due to a pair of beams (derived from the same laser) with relative detuning $\mu_R$. The beams cross with an angle of $\theta_R \simeq 5^o$ generating an optical dipole force with an effective wavelength of $\lambda_R =2\pi/|\vec{\Delta k}|\approx$ 3.7 $\mu$m.  }
\label{exp setup}
\end{figure}

In this section we discuss our experimental set-up for generating
Ising interactions between several hundred $^{9}$Be$^{+}$ ions localized
in single-plane triangular arrays confined by a Penning trap. Axial confinement
in the direction of the magnetic field ($\hat{z}$-direction, $B_0$=4.46 T) is due
to the electric field from a quadrupolar trap potential $\phi_{T}=\frac{1}{2}m_{Be}\omega_{z}^{2}\left(z^{2}-\rho^{2}/2\right)$,
where typically $\omega_{z}\simeq2\pi\times$795 kHz. Rotation at
frequency $\omega_{r}$ (about $\hat{z}$) produces a radial restoring
potential due to the velocity-dependent Lorentz force. $\omega_{r}$
is precisely controlled with an external rotating quadrupole potential
(a {}``rotating wall'')~\cite{Huang:1998}. For a weak rotating wall, the trapping
potential in the rotating frame is $q\phi(\rho,z)\simeq\frac{1}{2}m_{Be}\omega_{z}^{2}\left(z^{2}+\beta\rho^{2}\right)$,
where $\beta=\omega_{r}\omega_{z}^{-2}\left(\Omega_{c}-\omega_{r}\right)-\frac{1}{2}$.
For $100 \leq N \leq 350$, we set $\omega_{r}\approx2\pi\times45$
kHz so that the radial confinement is weak enough that a cloud of
ions relaxes into a single 2D plane $(\beta<<1)$. When the ions\textquoteright{}
motional degrees of freedom are Doppler laser cooled ($<1$ mK), the
ions naturally form a 2D Coulomb crystal on a triangular lattice,
which is the geometry that minimizes their mutual Coulomb
potential energy~\cite{Mitchell:1998}. See Fig.~\ref{exp setup}(a) for an image of the triangular lattice
taken in the rotating frame of the ions.

Reference~\cite{Biercuk:2009} gives a detailed discussion of our spin initialization,
control, and measurement capabilities with planar ion arrays in Penning
traps. Here we briefly summarize some of that discussion. Figure ~\ref{exp setup}(b)
shows the relevant $^{9}$Be$^{+}$ energy levels. We use the valence-electron spin states parallel $\left|\uparrow\right\rangle =\left|m_{J}=+\frac{1}{2}\right\rangle $
and antiparallel $\left|\downarrow\right\rangle =\left|m_{J}=-\frac{1}{2}\right\rangle $
to the applied magnetic field of the Penning trap as the two-level
system or spin. In the 4.46 T magnetic field of the trap, these levels
are split by approximately 124 GHz. Spins in the $\left|\downarrow\right\rangle $
state are efficiently optically pumped to the $\left|\uparrow\right\rangle $
state by a laser tuned to the $\left|\downarrow\right\rangle \rightarrow\left|2P_{3/2,\:}m_{J}=+\frac{1}{2}\right\rangle $
transition. The repump beam and the main Doppler laser cooling beam
are directed along the magnetic field ($\hat{z}$-axis). In addition,
a weak Doppler laser cooling beam (40 $\mu$m waist) directed
perpendicularly to the $\hat{z}$-axis directly Doppler cools the
perpendicular degrees of freedom. A typical experimental cycle starts
with 10 ms to 20 ms of combined Doppler laser cooling and repumping.
The fidelity of the $\left|\uparrow\right\rangle $ state preparation
is estimated to be very high (99.9 \%).

Low-phase-noise microwave radiation at 124 GHz is used to rotate the
spins. The microwave source consists of an agile 15.5 GHz source
followed by an amplifier multiplier chain with 150 mW output power
at 124 GHz. The microwaves are transported to the ions down the bore of the magnet with a rigid waveguide
and directed onto the ions with a horn located between the ring and
endcap electrodes of the trap. We obtain $\pi$-pulses (the duration
of time required to coherently drive $\left|\uparrow\right\rangle $
to $\left|\downarrow\right\rangle $) of 70 $\mu$s duration. The measured spin-echo coherence duration
($T_{2}$) is $\sim$100 ms.

At the end of an experimental sequence we turn on the Doppler cooling
laser and make a projective measurement of the ion spin state through
state-dependent resonance fluorescence. With the Doppler cooling laser
on, an ion in the $\left|\uparrow\right\rangle $ state scatters $\sim10^{7}$
photons/s, while an ion in $\left|\downarrow\right\rangle $ is dark.
For the spin precession measurements summarized here we performed
a global spin-state detection. Specifically we detected, with f/5
light collection optics and a photomultiplier tube, the resonance fluorescence
from all the ions in a direction perpendicular to the magnetic field.

We engineer an effective Ising interaction with a spin-dependent optical dipole force (ODF) obtained from off-resonant laser light. The ODF is directed either towards or away from regions
of high laser intensity, depending on the frequency of the laser light
relative to an atomic transition frequency. An optical dipole force
is the analog of the well known ponderomotive force in plasma physics,
here applied to the bound valence electron of the $^{9}$Be$^{+}$
ion. We employ a spin-dependent ODF generated
from the interference of two off-resonant laser beams at the ion cloud position, as depicted schematically in Fig.~\ref{exp setup}(c). If the frequencies of the
two laser beams are the same, a one-dimensional optical lattice (i.e.,
1D intensity standing wave) is generated at the intersection of the
two off-resonant laser beams. With a frequency difference $\mu_{R}$
between the two laser beams, the intensity standing wave becomes a
``moving wave'' with the intensity maxima passing through the
single ion plane at a frequency $\mu_{R}$. This generates an oscillating
optical dipole force $\propto\cos(\mu_R t)$. By adjusting the frequency and polarizations
of the laser beams, we generate forces that are equal and opposite
on the $\left|\uparrow\right\rangle ,\,\left|\downarrow\right\rangle $
spin states~\cite{Britton:2012},
resulting in the ODF interaction,
\begin{equation}
\hat{H}_{ODF}=-\sum_{j=1}^{N}F_{0}\cos\left(\mu_{R}t\right)\hat{z}_{j}\cdot\hat{\sigma}_{j}^{z}\;.\label{eq:ODF}
\end{equation}
Here $N$ is the total number of spins in the array, $\hat{z}_{j}$ is the axial position operator for ion $j$, and $\hat{\sigma}_{j}^{z}$ is the z-component of the Pauli spin matrix for ion $j$. A typical optical dipole force $F_{0}$ used in our work is $\sim2\cdot10^{-23}$
N.

As discussed in the previous section, $\hat{H}_{ODF}$ can generate an effective spin-spin interaction. Quantum
mechanically, one solves for the evolution operator associated with
the Hamiltonian $\hat{H}_{ODF}$. Because $\hat{H}_{ODF}$ explicitly
depends on time, this could involve an infinite sequence of commutators
of $\hat{H}_{ODF}$ with itself at different times. However, this
sequence truncates after two terms. One term involves a product of $\hat{z}_{j}$ and $\hat{\sigma}_{j}^{z}$,
and describes spin-motion entanglement generated by $\hat{H}_{ODF}$~\cite{Sawyer:2012}.
This term can be minimized by adiabatically turning on and off the
interaction, as discussed in the simple examples in the previous section.
The second term, which we label $\hat{H}_{I}$, involves the
product of different spin operators and describes the effective spin-spin
interaction~\cite{Kim:2009},
\begin{equation}
\hat{H}_{I}=\frac{1}{N}\sum_{i<j}^{N}J_{i,j}\hat{\sigma}_{i}^{z}\hat{\sigma}_{j}^{z}\;.\label{eq:H_I}
\end{equation}
Here $J_{i,j}$ is the strength of the effective Ising interaction between ion (or
spin) $i$ and ion (or spin) $j$. The $J_{i,j}$'s
can be expressed in terms of the transverse (or drumhead) modes $\left(\omega_{m},\vec{b}_{m}\right)$
of the array, where $\omega_{m}$ is the mode eigenfrequency and $b_{j,m}$
(the jth element of $\vec{b}_{m}$) denotes the relative amplitude
of ion $j$,
\begin{equation}
J_{i,j}\simeq\frac{F_{0}^{2}N}{2\hbar m_{Be}}\sum_{m=1}^{N}\frac{b_{i,m}b_{j,m}}{\mu_{R}^{2}-\omega_{m}^{2}}\;.\label{eq:Jij}
\end{equation}

\begin{figure}
  \includegraphics[height=.3\textheight]{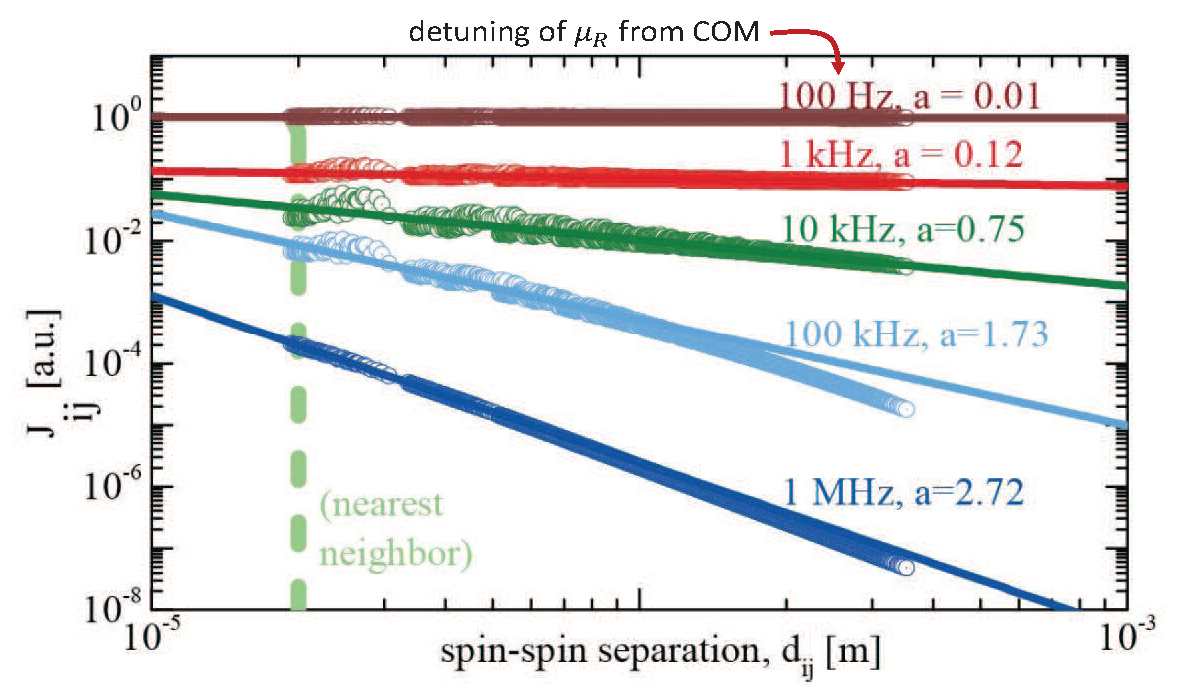}
  \caption{Explicit calculation of the pairwise coupling coefficients $J_{i,j}$ for
$N=217$ spins and plot as a function of spin-spin separation, $d_{i,j}$. For
$\mu_R-\omega_z < 1$ kHz, $\hat{H}_{ODF}$ principally excites COM motion in which all ions
equally participate: the spin-spin interaction is spatially uniform. As the detuning
is increased, modes of higher spatial frequency participate in the interaction and $J_{i,j}$
develops a finite interaction length.}
\label{power law}
\end{figure}

Equation \ref{eq:Jij}, along with a calculation of the drumhead modes,
can be used to determine the effective spin-spin interactions. Figure~\ref{power law} shows the results of such a calculation for different positive
detunings of the optical dipole force beatnote $\mu_{R}$ from the
center-of-mass (COM) mode $\omega_{z}$. Note that the dispersion of the
drumhead modes is anomalous; the longest-wavelength mode $(\omega_{z})$
is the highest-frequency drumhead mode. Figure~\ref{power law} shows that for $\mu_{R}>\omega_{z}$,
the pairwise coupling coefficients are positive $\left(J_{i,j}>0\right)$,
resulting in an antiferromagnetic interaction, in agreement with the
simple arguments of the previous section. As the detuning $\mu_{R}-\omega_{z}$
is increased, the overall interaction strength weakens, and the
range of the interaction gets shorter. In general the range of the
interaction can be characterized by an approximate power law $J_{i,j}\sim J/\left|\vec{r}_{i}-\vec{r}_{j}\right|^{a}$,
where $a$ can be tuned between 0 and 3. The spin-spin
interactions readily generated with trapped ions are long range. These
long-range spin-spin interactions are ``unphysical''
in the sense that the strongest spin-spin interactions in solid-state
materials are due to the exchange interaction, and therefore nearest
neighbor. However, long-range interactions provide a good mechanism
for generating long-range quantum entanglement, which appears to be a property of states with nontrivial topological/quantum
order~\cite{Chen:2010}. Therefore the long-range interactions readily obtained
with trapped ions provide an interesting contrast with naturally occurring
solid-state systems, and an interesting place to search for new emergent
behavior.

\begin{figure}
  \includegraphics[height=.41\textheight]{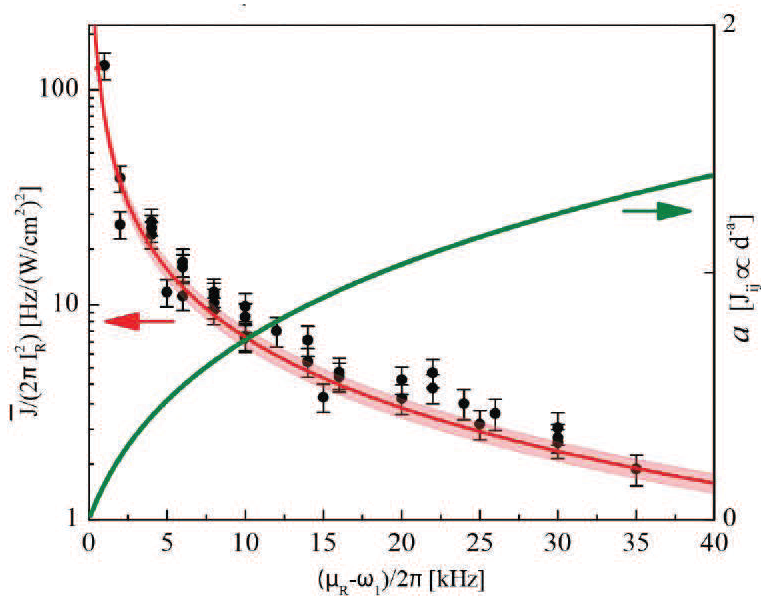}
  \caption{Measurements of $\bar{J}$, normalized to the measured laser intensity, as a function the ODF beatnote detuning from the COM mode. The solid line (red) is the prediction of mean-field
theory that accounts for couplings to all $N$ transverse modes; there are no free
parameters. The line's breadth reflects experimental uncertainty in the angle
$\theta_R=4.8\pm0.25^o$. The mean-field prediction for the average value of the power-law
exponent, $a$, is drawn in green (right axis, linear scale). }
\label{benchmarking}
\end{figure}

The assumptions utilized in deriving Eq. \ref{eq:Jij} are substantial.
For example, we assume that each ion satisfies a Lamb-Dicke confinement
criterion~\cite{Britton:2012} based on the ``wavelength'' of the optical dipole force,
and that the optical dipole force wavefronts are precisely aligned
with the ion planar array, to name just a few. It is important to benchmark the experimental system to assure that it behaves like Eqs. \ref{eq:H_I} and \ref{eq:Jij}. We do this by comparing
experimental data with the mean-field prediction that the influence
of $\hat{H}_{I}$ on spin $j$ can be modeled as a magnetic field
$\bar{B}_{j}=\frac{2}{N}\sum_{i,i\neq j}^{N}J_{i,j}\left\langle \hat{\sigma}_{i}^{z}\right\rangle $
in the $\hat{z}$ direction due to the remaining $N-1$ spins. The
details are discussed in Ref.~\cite{Britton:2012}
and supplementary material. This interaction-induced magnetic field
gives rise to spin precession in excess of normal Larmor precession.
We prepare all the spins in the same initial state, rotated by an
angle $\theta$ from the $\hat{z}$ axis (i.e., prepare spin $\left|\uparrow\right\rangle$ and then rotate by $\theta$), and measure the excess spin precession as a function of $\theta$.
Through global fluorescence measurements we measure the excess precession
averaged over all ions, from which we are able to extract
the average pairwise coupling $\bar{J}=\frac{1}{N^{2}}\sum_{j=1}^{N}\sum_{i,i\neq j}^{N}J_{i,j}$.
Figure~\ref{benchmarking} summarizes measurements of $\bar{J}$
as a function of the beatnote detuning $\mu_{R}-\omega_{m}$. With
no fitted parameters these measurements agree well with the predictions
of Eq. \ref{eq:Jij}, including contributions from all modes. Also
shown in Fig.~\ref{benchmarking} is a theoretical calculation of the power-law exponent,
which describes the range of the interaction. At the mean-field level,
our system is well described by Eqs. \ref{eq:H_I} and \ref{eq:Jij}.

Future efforts include benchmarking quantum effects such as depolarization~\cite{Kastner:2011} and squeezing/anti-squeezing due to the quantum many-body interaction.  An intractable simulation requires adding a non-commuting term to $\hat{H}_I$, for example a transverse magnetic field, which we can do with the 124 GHz microwaves. Several technical improvements likely need to be made before this step.  For example, the coherence of $\hat{H}_I$ is currently limited by spontaneous emission from the off-resonance laser beams.  This can be greatly reduced by increasing the angle $\theta_R$ between the two ODF laser beams.


\begin{theacknowledgments}
  Work supported by the DARPA OLE program and NIST.  We acknowledge useful discussions with M. Biercuk and H. Uys. This manuscript is a contribution of the US National Institute of Standards and Technology and is not subject to US copyright.
\end{theacknowledgments}



\bibliographystyle{aipproc}   



\end{document}